\documentclass[]{interact}

\usepackage{url}
\usepackage{float}
\usepackage{epstopdf}
\usepackage[caption=false]{subfig}

\usepackage[numbers,sort&compress]{natbib}
\bibpunct[, ]{[}{]}{,}{n}{,}{,}
\renewcommand\bibfont{\fontsize{10}{12}\selectfont}
\makeatletter
\def\NAT@def@citea{\def\@citea{\NAT@separator}}
\makeatother

\theoremstyle{plain}

\theoremstyle{definition}

\theoremstyle{remark}

\begin{document}

\articletype{}

\title{Encryption by using base-n systems with many characters}

\author{
\name{A. Hoenen\thanks{CONTACT A. Hoenen. Email: hoenenarmin@gmail.com}}
\affil{Institut f\"{u}r Empirische Sprachwissenschaft, Goethe Universit\"{a}t, Frankfurt, Germany}
}

\maketitle

\begin{abstract}
It is possible to interpret text as numbers (and vice versa) if one interpret letters and other characters as digits and assume that they have an inherent immutable ordering. This is demonstrated by the conventional digit set of the hexadecimal system of number coding, where the letters ABCDEF in this exact alphabetic sequence stand each for a digit and thus a numerical value. In this article, we consequently elaborate this thought and include all symbols and the standard ordering of the unicode standard for digital character coding. We  show how this can be used to form digit sets of different sizes and how subsequent simple conversion between bases can result in encryption mimicking results of wrong encoding and accidental noise. Unfortunately, because of encoding peculiarities, switching bases to a higher one does not necessarily result in efficient disk space compression automatically.
\end{abstract}


\section{Introduction}

One can hide a message within another message. To this end hiding text transmitting numbers as ciphertexts corresponding to a non-numerical plaintext or vice versa is an age old concept. Natural languages express words in the written modality through a conventionalized set of characters and conventionalized mappings from the acoustic modality to the visual one -- through a writing system with an orthography. Numbers are usually expressed using a likewise conventionalized set of digits with an immutable ordering cinstituting the mapping to idealized and progressing ($+1$) integer values starting in $0$. One arranges them in a rigorously defined placement system, where each position represents a base which exactly corresponds to the size of the digit inventory raised to a certain power, progressing from right to left, starting in $0$ as the lowest power. We arrived at such a system with the default inventory of $10$ digits including $0$ only relatively recently, but it has since become an international standard especially for scientific interchange and a largely language independent form of mathematical expression in daily life. 
These givens entail a clear parallelism between numbers and texts: we use conventionalized sets of symbols which have an inherent ordering within the set (integer progression for numbers, the alphabet for letters) and arrange them into basically possibly infinit sequences. There are differences, two of the most important of which are: Firstly, the size of the set with the decimal digit set for integers containing only 10 symbols, whereas the alphabet contains 26 letters and is the minimal set for representing language. This entails under more a much larger combinatorial space for letters and a larger possible entropy. Secondly, the rules for arrangement are quite rigorous in math, but less so for letters, where we can arrange them according to our communicative needs in almost any sequence (think about abbreviations, acronymes, all kinds of word- and letterplay, codes etc.). These differences can and must be mitigated if one wants to exploit the parallelism using letter symbols as mathematical digit sets for instance in order to produce or use interpretational ambiguity. The mathematical inventory needs thus to be enlarged character sets and those must obey or map to the stringent positioning rules of mathematics. This means, that the otherwise not extremely strictly necessary inherent ordering represented by the somehow arbitrary alphabetical sequence must, in case of usage of characters as digits, be absolutely fixed. Both of this can be done easily. In math, we can use a larger inventory by swapping the numerical base of a system and for characters (and digits), we can use the technically fixed and globally adopted standard ordering of symbols from the unicode standard.

Apart from this, textual characters forming written expression and digits forming numbers are correlated in many meaningful ways. Nowadays, converting text into numbers and computing its properties is one of the foundations of natural language processing the subdiscipline of computer science closely related to statistics of language and with this only a small step away from both mathematics and cryptoanalysis. 

\section{History}

Numeral systems with bases other than $10$ have been frequent in history. In Mesopotamia, a mixed system with a large importance of the number of $60$ (therefore sometimes referred to as \textit{sexagesimal}) was in use, compare \cite[p.247]{chr10}.
In ancient China, a base-16 system was used for measurements and other examples are extant. Furthermore, roman numerals, widely in use in Europe until early modern times used letters as digits, which was also quite common in some other writing systems such as Hebrew, Arabic etc. In the latter, letters were used with certain numerical values assigned to them, sometimes combined with positioning or other contextual clues. This already resulted in some encipherment using the ambiguity of interpretation between words and numbers. Consider \cite[p.66]{lang21}:

\begin{quote}
\noindent Al-Qalqashandi mentions a procedure in which two Arabic letters correspond to a letter of the plaintext in such a way that the numerical values of the two letters equal the numerical value of the plaintext character
\end{quote}

\begin{figure}
\centering
\label{fig:1}
\includegraphics[scale=0.5]{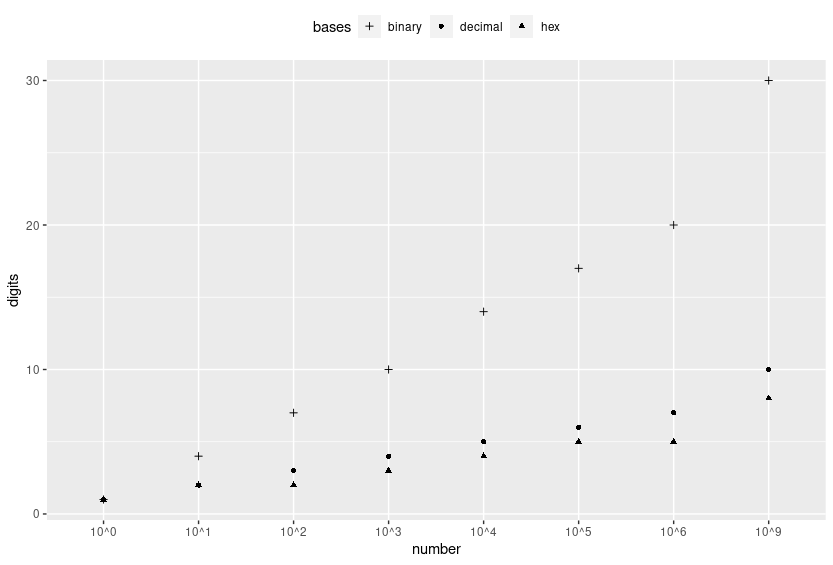}
\caption{Lengths in number of required digits/places (y-axis) for some low powers of $10$ (x-axis) in the binary, decimal and hexadecimal systems.}
\end{figure}

\noindent However, in the modern age, with the advent of computers the binary and hexadecimal systems became important. Two digits $0$ and $1$ -- simplistically speaking -- represent $2$ states of electric current. Unfortunately, with a base-2 system even smaller numbers become very long very quickly.
The hexadecimal system to the contrary represents numbers with even fewer places than the decimal system and has the advantage to be easily convertible to and from the binary system as it is based on a low power of $2$: $2^4$. Figure \ref{fig:1} shows how large the difference in length is for some low powers of $10$ depending on the system. As one can see, the larger a number gets, the more pronounced is the length difference between lower and higher base-systems for encoding the same number.

\section{Terminology}
Terminologically, \textit{number system} often refers to the kind of numbers we use. This means primarily large sets from number theory, such as $\mathbb{N}$ standing for \textit{natural numbers}, $\mathbb{R}$ for real numbers and so forth. A \textit{numeral system} on the other hand refers to a set of symbols used for expressing numbers. The term itself does include all kinds of differing symbol systems such as Sumerian numerals, Cistercian numerals, Roman numerals and so forth including also our Western Arabic numerals. However, this does not specify anything about a positional use. Roman numerals for instance were arranged according to completely different rules than our modern numerals and the base of the system could not even be determined from the size of the inventory. When we use a standard positional system, we may call it a positional numeral system or the standard positional numeral system and within this system, we can distinguish different bases and name subsystems base-n, with n most commonly a positive integer. However, such terminological conventions are not completely strict and one finds the use of the term \textit{number system} likewise for base-n systems not necessarily entailing their use as actual sets in the sense of number theory, compare for instance  \cite{aki08}. Also, one finds even more terms such as \textit{numeration system}, \textit{counting system} and others more used interchangeably to refer to number sets, numeral systems or positional numeral systems or yet other entities. The exact meaning of the terms used should always be clear from the context. Within this article, we will use \textit{numeral} and \textit{digit} interchangeably for a symbol in a mathematical system, \textit{number system} for sets from number theory and \textit{numeral system}, \textit{(standard) positional numeral system}, ... for our conventional decimal system and/or mentioning the base for base-n systems. Quotes and titles in the references may not adhere to that usage.

\section{Mathematical wrap-up on numeral systems}
There has been plenty of research on (positional) numeral systems in mathematics, computer science and cryptography/cryptoanalytics. In mathematics, research continues until today. One important branch is concerned with different qualities of the bases: negative bases \cite{grun85}, complex numbers as bases \cite{knuth60}, rational numbers as bases \cite{aki08} or particular bases, such as $-\frac{3}{2}$ as in \cite{ross22}. In cryptography exponentiation methods and their effective implementation play a crucial role, compare \cite{gord98}. The interested reader may be referred to these examples and the circumspanning body of literature for flanking information. The aim and method of this paper are much simpler as will be explained in the next section. However, whilst systems with larger bases are being discussed in the aforementioned literature, they are mostly expressed as formula, rather than spelled out as actual numbers, since for the sake of the argument this very often is not necessary. 
Generally, any number $x$ in our standard positional numeral system can be expressed as:
\begin{equation}
x = \sum_{n=0}^{m}{a_nb^n}
\end{equation}
where $a$ is a digit from the ordered symbol inventory of the particular numeral system\footnote{Here, we limit ourselves to looking at numeral systems with a $b$ from $\mathbb{N}^{+}$.} corresponding to a value (for instance $B$ in the hexadecimal system has the value $11$), $n$ is the position within the number starting to the right indexed as $0$ and $b$ is the given base. $m$ is the leftmost positional index, which, since we start counting in $0$ equals $pl(x,b)-1$ with $pl(x,b)$ the function returning the number of positions or length of a number given a base. At each position, each of the symbols of the digit inventory (numeral) may appear (apart from leading zeroes) and thus $a$ can take any value from $0$ to $b-1$ with the exception of the leftmost position where $0$ is disallowed for convenience.
As an example, the hexadecimal number $ABC$ can be expressed as $12 \cdot 16^0+11 \cdot 16^1 + 10 \cdot 16^2$. The base $b$ always corresponds to the size of the symbol inventory whilst the largest value expressed per position for $a$ is $b-1$. 

As one can see, for this notation, one does not need symbols other than those used in the decimal system plus some standard mathematical symbols ($\cdot,+$ and superscripting). So, we can express any hexadecimal $x$ as a decimal sum: $12 \cdot 16^0+11 \cdot 16^1 + 10 \cdot 16^2$. Extending this to the conventionalized set of mathematical symbols such as $\sum$, we can express basically any number or entire systems omitting the use of other than the decimal inventories digits. Hexadecimal $ABC$ could be expressed as:
\begin{equation}
\sum_{n=0}^{|A|-1}{a_nb^n}; b=16; A=\{12,11,10\} 
\end{equation}
or even more simplistically use something like $10-11-12$. This would make such a number readable\footnote{The reader may do a thought experiment and ask herself how quickly she could map $A$, $B$ and $C$ to $10$, $11$ and $12$ automatically.} and one would not have to bother with choosing or printing a set with an inherent ordering. 
In general, spelling out such numbers in other less used numeral systems is rather rare and there is good reason for it. We are not used to write and spell out such numbers apart from some few exceptions such as the binary and hexadecimal systems and furthermore, we would hardly be able to read and decode such information. Finally, a conventionalized sequence for the digit inventories should be pre-established and how to do that and which symbols to use is to the best of the authors knowledge, not agreed upon or standardized in the broader field of mathematics. 

However, in the field of computer science, this is different as will be seen later. For the representation of numbers in the hexadecimal (and binary) system(s) where the digit inventory is conventionally fixed (so one would not need to print the inventory and sequence for decoding when reading), the need sometimes arises to write the basis, especially because otherwise some ambiguities arise. Each number of the decimal system is also a number in the hexadecimal system and each binary number is also one in the decimal and the hexadecimal systems, since the digit inventories are in ordered subset relations. Whilst $0$ remains zero and $1$ remains the value one in each of the systems due to being the first digits in the ordered inventory, the sequence $10$ already has a different value depending on the system. In the binary system, it is $2$, in the hexadecimal system, it is $16$ and, interestingly, in any system with a digit inventory that starts in the $10$ arabic numeral digits, it has the value of the base of the system. In order to clarify which meaning is intended, we can write $10_{b},10_{d},10_{h}$ or $(10)_{2},(10)_{10},(10)_{16}$. We will use the notation of parenthesis and subscript in decimal since it would be unclear/overcomplex to verbally convert any n-base into a greco-roman number word and then find some abbreviation, or in other words, whilst the \{b,d,h\}-notation avoids reading ambiguity subscripting a decimal number, it is only really usable for binary, decimal and hexadecimal systems.

\section{Spelling out base-n}
The problem of spelling out a system with a base larger than $10$ resp. $16$ is the first focus of this paper. To this end, the hexadecimal system was one of the first, where the concrete application led scholars and practicioners to experiment and finally converge on a standard for the representation of the digits representing the values/numbers $10$ through $15$. For the binary system, $0$ and $1$ as a subset of the digit inventory of the decimal system have been established, but it is of course easier to build subsets from an already existing set than to agree on how to expand a set.
At first, different groups used different symbols.\footnote{For reference see \url{https://en.wikipedia.org/wiki/Hexadecimal}, subsection "Other symbols for 10–15 and mostly different symbol sets", version from Jun 4, 2023.}

There were even scholars inventing new naming/pronunciation conventions, for instance \cite{mag68}. It would of course greatly improve readability, had we special words for digits representing values in other than the decimal system, but it seems that by and large for an average human being one numeral system in the natural languages we talk is already enough. The verbal expression of numbers is again a linguistically complex phenomenon with all kinds of exceptions and complexities such as the (in)famous number $80$, expressed as $4$ (times) $20$ in standard French and e.g. Basque. Adding to this complexity by inventing and teaching new names for digits of the hexadecimal system remained a curiosity.
Apart from the hexadecimal and binary systems, other systems have been elaborated on albeit not as rigorously as the former two, which are used more often also for mathematical argument.

The second focus of this paper is how to use the first system in connection with de- and encipherment and compression. With the now quite conventionalized sequence of (uppercase or lowercase, but not mixed case) A-F, coincidentally some hexadecimal numbers can form words of English such as $(BEE)_{16}$ which corresponds to $(3054)_{10}$. Whilst in the hexadecimal system, this is a rare coincidence, one can systematically devise a system which does interpret all text as digits of a base-n system.

\section{Method}
The easiest base-n system for English would be a base-36 system, where the 10 digits plus the 26 letters of the alphabet in alphabetic sequence would be in use. This system would just append all other letters after F to the set used by the hexadecimal system, exactly in the alphabetic sequence. Let us suppose we take only lowercase letters and no punctuation marks. Each word would then express a numerical value. However, not every possible digit sequence in that system would correspond to an actual word. 

\begin{figure}
\label{fig:2}
\textbf{1 - Original text}: This is an example. \\
\textbf{2 - Computing the conversion, example}: this \\ 
$28 + 18 \cdot 36+17 \cdot 36^2+ 29 \cdot 36^3$,\\ starting in 28 since s is the 29\textsuperscript{th} digit in the ordered set\\ $\{0,1,2,3,4,5,6,7,8,9,a,b,c,d,e,f,g,h,i,j,k,l,m,n,o,p,q,r,s,t,u,v,w,x,y,z\}$ \\ and stands thus for value $28$ as we start in $0$. For computation, we start from the rightmost digit and multiply its value with $36^0$, then we proceed with i analogously standing for $18$ which we multiply with $36^1$ and add to the previous result and so forth.\\
\textbf{3 - Text converted to base-10}: 1 375 732, 676, 383, 32 488 192 274 \\
\textbf{4 - Simple ciphertext with punctuation marks, X for capital letter, and a hyphen as separator inserted}: 
X1375732-676-383-32488192274.
\caption{A simple encryption based on interpreting text as numbers in a base-36 number system, which extends the hexadecimal digit set to the entire alphabet.}
\end{figure}

The result of a conversion can be seen in Figure \ref{fig:2}. In itself, such a conversion could already be used as a ciphertext, especially for short messages and when some ad-hoc modifications add idiosyncratic semantics as in step 4. Such a cipher for longer texts could be hard to spot in the context of number code files, but if spotted easy to break. For instance the number of words per sentence is quite obvious and the statistics of tokens per sentence per language could be a first approach identifying such a ciphertext as encoding of natural language and a hint towards the underlying language itself. Token frequency statistics would quickly open the road to decipherment.
Another problem of this text is its length. While the original input has 19 characters including spaces, the idiosyncratic encoding has 29. These problems come under more from the fact that 36 is a higher basis than is 10 and converting to lower bases usually inflates length. A work-around for the first problem would be to enhance the system to encode space as a character making it a somewhat unhandier base-37 system. Also the sequential position of space could be debatable and important for decipherment: would we want to put it to the beginning before 0 with the funny but not so unwelcome effect of even shifting actual numbers in the plaintext, even 0 and 1. Or would one put space between digits and letters or to the end? For the sake of exemplification, we put it to the end as last symbol in the set. One could also include punctuation. In order to disguise the number and frequency statistics of tokens now, one could chunk the text into equally sized bits. This could also benefit length issues for very long words which otherwise would lead to extremely long base-10 encodings. So, if we chose 5 as chunk-size we would arrive at the scenario in figure \ref{fig:3}. 

\begin{figure}
\label{fig:3}
\textbf{1 - Original text}: this is an example.\\
\textbf{2 - Chunked}: (this ),(is an),( exam),(ple).\\
\textbf{3 - Conversion}\\
\textbf{4 - Text converted to base-10}: 55237484-35202859-68224507-35016.\\
\caption{A slightly more sophisticated conversion with an additional step: chunking.}
\end{figure}

The new encoding just incidentally has as many numbers as words in the plaintext. Obviously, one could also use hex-encoding instead of decimal encoding, making the sequence mimick some computer relevant machine readable hex sequence of some kind. One would however rely on not being detected. One weakness is the smaller size of the last number, since only the last chunk can but must not become shorter than 5 or any chosen chunk size. Obviously, one could use padding and fill the last word until it too had 5 positions: (pleXX) and simply ignore that in the decipherment. The decipherment itself would still be rather simple if one knew about the method. Knowing the chunk size would not even be necessary, only knowing the base of the system and with rather conventionalized sequences for all letters of the English alphabet and even adding space and some punctuation marks brute-forcing all possible base systems with their inventories would not yet represent a problem for modern computers.

An obvious further ingredient to encipherment could be to use any sequence determining chunk sizes. Imaginable would be for instance any integer sequence of the OEIS\footnote{\url{https://oeis.org/}}. This could be transmitted as a key. In our example, one could chunk according to the prime numbers, with the then key 40 (or A000040).\footnote{\url{https://oeis.org/A000040}} One could prefix or suffix this number to the ciphertext or even add it to the first number and only the intended receiver would know that. Of course, periodical sequences with not too large numbers would fit the purpose best in order to avoid length issues. This however could already be one vulnerability. All kinds of additional play could be applied such as always subtracting one, submitting a second equally long sequence of different addends which have been added to the numbers of the first sequence and so forth. An example can be seen in Figure \ref{fig:4}. Also one could chain such encodings.

Whilst such plays could add to the safety of the method, basically it would not alter the principal vulnerability caused by the rather small size of the original inventory and the concurrently rather limited number of possible conventionalised sequences one would have to try. Even if the code-breaker would not know exactly what size the inventory had and whether to include punctuation, diacritic marks etc. the number of possible ordered sets would still be small. When considering the possibility of different languages and base alphabets, things already become more sophisticated, but still the number of writing systems in the world is limited and for sure one would have an idea of which languages or at least writing systems could be relevant. Again playful variants writing English with Cyrillic letters etc. would exist but not basically change the game.
Also, since hexadecimals are used in various machine codes, one could mimick such files or use channels such as internet packages to transmit them.

\begin{figure}[H]
\label{fig:4}
\textbf{1 - Original text}: this is an example.\\
\textbf{2 - Chunked}: (th), (is ), (is an), ( exampl), (e)\\
\textbf{3 - Conversion}\\
\textbf{4 - Text converted to base-10}: 1130-25714-35202859-3,455775984$\cdot 10^{12}$-14\\
\textbf{4a - Text converted to base-16}: 46A-6472-219272B-BB56869BF64DFF0000-E\\
\caption{A yet slightly more sophisticated conversion. Here with idiosnycratically using a power of ten notation for a large number.}
\end{figure}

Some readers may have already felt reminded of a similar encoding being in use. Namely as the Base64 standard encoding for 8-bit binary files. Here, the conventionalized sequence includes 26 uppercase letters, 26 lowercase letters, 10 digits and two additional characters: $/$ and $+$ amounting to 64 ($2^6$) characters ordered as uppercase-alphabet, lowercase-alphabet, digits, additional characters, thus the ordering differs from the hexadecimal system. The two additional characters have been chosen rather arbitrarily and cause some problems for instance for filenames. They were needed in order to have exactly 64 characters and they had to be from the ASCII standard base set, which did not provide other letters. 
Besides Base64, some other bases have been established in the world of computers. RFC 4648 \cite{rfc06} is testimony to the fact that standardization is attempted. It lists the Base64 alphabet, but also states about set and sequence, p.5:
\begin{quotation}
\noindent There is no universally accepted alphabet that fulfills all the requirements.
\end{quotation}
\noindent This RFC referring also to previous systems, defines the hexadecimal system inventory and sequence using upper case letters.
We also find some observation in relation to attacks on such encodings, p.14:

\begin{quotation}
\noindent Base encoding adds no entropy to the plaintext, but it does increase the amount of plaintext available and provide a signature for
cryptanalysis in the form of a characteristic probability distribution.
\end{quotation}

\noindent Basically, if one were to permute the digits and all letters of the alphabet to an idiosyncratic sequence just for the purpose of encipherment, then brute-forcing could be made much more difficult, since combinatorically one can produce $n!$ different ordered sequences or permutations, for $n=37$ symbols thus $37!$ sequences, a huge number with $44$ decimal places. If one now also has to decide which exact subset of characters is to be chosen (the encoder could easily choose characters which do not appear in the plaintext or restrict himself to those) and how large it is, the number of possibilities becomes too large to handle. However, the sequence and inventory would have to be known to the receiver which is a new vulnerability, especially without such sophisticated mechanism as private and public keys to transmit this. One virtue of the non-permuted sets lies in the fact that the sequence is near-conventionalized and thus can be known without transmission to the receiver. If one wanted to leverage the power of combinatorics still and without permutation, then there would be yet another way to improve on difficulty. That of expanding the set to all encodable characters and using the conventionalized sequence of the unicode standard. 

This would mean having a base-n system, where all letters of the languages of the world -- or all which can occur in any document -- are included. n must be sufficiently large so as to include all letters of the alphabet. Those letters must have a fixed sequence allowing them to be mapped to the same digit all the time. The latter requirement is already met by the fact that alphabets usually have an inherent sequence, which amongst other things serves for learning them at an early age (think of the alphabet song). 

When it comes to alphabets such an inherent ordering is usually given. But already considering other languages than English, it might become an issue of how to order additional letters or diacritic symbols such as accents. If we take German for instance, the letters \"{a},\"{o},\"{u} and \ss do appear. They do have a conventional sequence which is widely acknowledged although there is still some variation in how dictionaries order entries with some letting \"{a}-words follow a-words whereas mostly they are appended to the alphabet, thus follow z-words.
When we now want to include other writing systems in order to convert multilingual text into numbers, the issue of a fixed sequence can at some point not be handled anymore by only choosing one sequence for each writing system but one would also have to choose the sequence of writing systems. 
The problem would also be another: for some writing systems with very large character inventories such as Chinese, a conventionally fixed order is not available.\footnote{Chinese characters in dictionaries are often ordered either by number of strokes or graphic elements which are called \textit{radicals} which then in turn are ordered by stroke number and for symbols with the same stroke number there are conventionalized sequences of the few basic strokes which make up all Chinese characters. Also stroke sequence is deterministic for each character.} To solve these ordering problems, one can use the Unicode standard.\footnote{https://unicode.org/}
This standard is administered by a consortium and publishes THE conventionalized set of (all) conventionalized, that is widely usable digital characters since 1991. It covers many scripts including character-heavy east asian ones and historical ones. Accidentally it also provides a sequence for all characters present. Although this sequence includes alphabetic sequences, it is largely an arbitrary ordering, which is irrelevant for our purpose. It provides a nearly perfect solution to our base-n problem.

We are furnished with a massive number of fixed sequence symbols. In fact, there are meanwhile more than 100 000 so called code points. A code point as basic unit of the standard can be a control character, in which case there is no visual equivalent. Furthermore, there are resrved areas and other special characters, as well as sections defining combined characters.
For a functioning conversion, one can exclude problematic characters. A sequence for a base encoding however must have each position filled. Thus excluding any character means that the following one will occupy the position and value subsequent to the one previous of the excluded character.
Ignoring this for a moment, we can prinicipally choose any n up to 100 000 or the current Unicode size and use this set as inventory with a large base and convert text into numbers for any of these systems.
All we have to do in order to convert/encode our text is the following.

\begin{figure}
\label{fig:5}
\begin{enumerate}
\item extract each code point present in the text
\item choose the highest codepoint number or one superior as max number of our base system
\item choose a preferably larger number as new base
\item extract the text from the file
\item escape linebreaks to some longer idiosyncratic sequence
\item choose a chunksize
\item chunk the text into slices of the chosen chunksize
\item for each slice convert from base system to new larger base
\item write to file
 \end{enumerate}
\caption{Simple succession of steps necessary to encode a text with any Unicode character into a higher base.}
\end{figure}

\begin{figure}
\label{fig:6}
\centering
\includegraphics[scale=0.3]{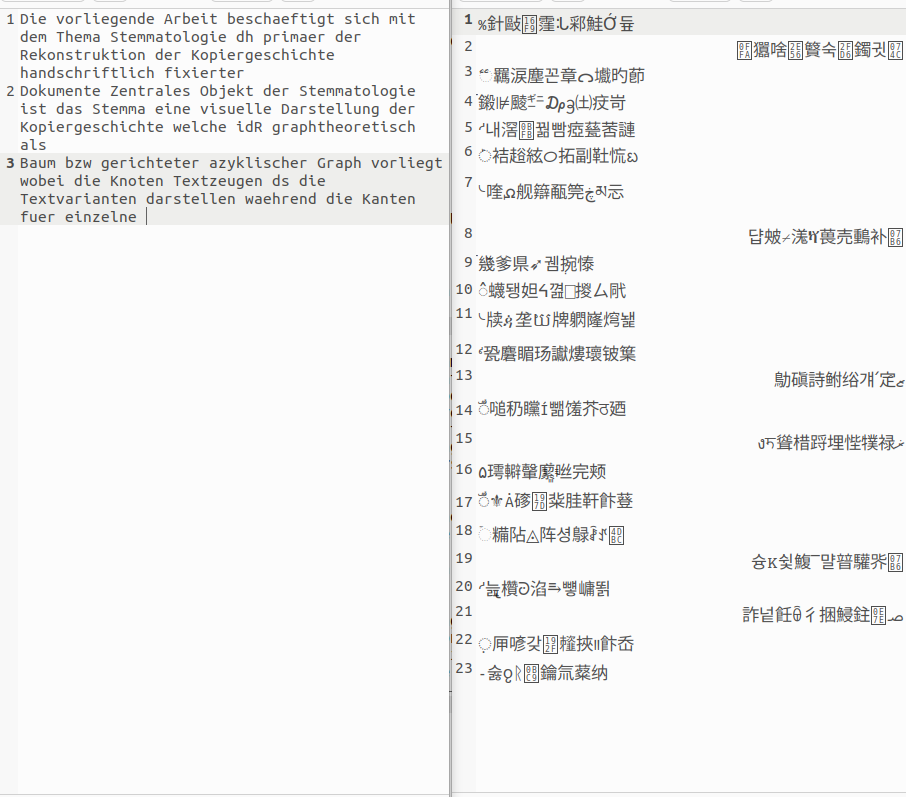}
\caption{Encoding of a simple text from base 200 to base 50 000. Note, that due to the encoding including control characters which determine writing direction, the output appears in lines with beginnings to the right and the left.}
\end{figure}

We implemented this in the Java programming language and Figure \ref{fig:6} shows a conversion example.

For decoding one simply follows the reverse way, but should know the encoding and decoding base. Of course, the algorithm allows for millions of variations and points at which a concealment strategy can be refined.

\section{Discussion}
The majority of characters produced are Chinese characters due to their ubiquity in the Unicode standard. The ciphertext may still appear similar to files where a wrong encoding has been applied, a wrong escaping, some filetype such as a binary interpreted as plaintext, maybe wrong OCR or broken output of some buggy computer program, or a purposefully generated random gibberish, a combination of this and so forth. The digital age will always continue to produce such or similar sequences naturally and this algorithm is probably one more potential addition to the noise production currently challenging cleaning methods for developers of machine learning datasets.

Instead of the purpose of concealment, where we can choose either a higher or a lower base in respect to the original, there is generally another aim why base system shifts to larger bases are being performed, namely compression. To this end, there exist independently developed similar implementations such as the ones propagated on github, which is not surprising given that such a use of unicode is an obvious choice and just enhances the hexadecimal set consistently.\footnote{\url{https://github.com/qntm/base65536},\url{https://github.com/qntm/base2048}} The authors mention compression ratios on their github repository and show, that such an encoding can be efficient for non variable byte-size schemes such as UTF-32. However, details are not available. Their use case is a compression of more information into a character limited Tweet rather than compression. 

Own tests suggested that the compression ratio due to the large disk space requirements of higher codepoints such as Chinese characters is not as effective as with known algorithms such as zip. The reason is that UTF-8 uses fewer bits than 8 if a character comes for instance from the ASCII part of the Unicode standard. Exploiting that standard for compression and more efficient information transmission, scholars have developed base64, see \cite{rfc06}. Similarly base85 is used and conventionalized for IPv6, see \cite{rfc1924}.  
A good overview is provided in the wikipedia.\footnote{\url{https://en.wikipedia.org/wiki/Binary-to-text_encoding}, subsection Encoding standards, last accessed Jun 4, 2023.} They became widely used for their technical benefit.


There is yet another peculiar base system conversion which seems to have come into being in relation with cryptography. It uses base-256 but no digit or character inventory, but instead uses a wordlist, the so-called PGP word list (from Pretty Good Privacy).\footnote{\url{http://web.mit.edu/network/pgpfone/manual/index.html#PGP000062}}
This particular encoding does not aim at compression as much, but instead of single characters/digits, words are used in a conventionalized sequence. This is very strongly reminiscent of the code books used in the cryptology of previous centuries. However again, this list as `key' does not have to be exchanged alongside the ciphertext as it is conventionally known. Base256 encoding produces a code, which can be pronounced and thus, for instance by a text to speech engine converted into the audio modality and vice versa as to date for English pretty good systems exist, such as openAIs Whisper \cite{whisp}.\footnote{\url{https://github.com/openai/whisper}}
Another special feature of the encoding is that depending on whether the target word stands at an even or odd position in the ciphertext, a different word is chosen.

These different base systems with larger bases have come into being and enrich or complicate the world of conveying messages. It seems however for such a system to establish, a strong usecase such as compression is needed. To this end, an encoding through the unicode set 

\section{Conclusion}
We presented a method to reencode a plaintext by interpreting it as a number in a sufficiently large base-n system including at least all characters of the text and then chunking and reencoding it in a larger base. As a conventionalized ordered inventory, we used the unicode standard. Encryption methods especially since private-key architectures are more sophistcated and more secure than the herepresented ideas presumably even if a random permutation of the available character inventories and some other idiosyncratic properties are chosen as has been demonstrated by small usecases and discussed. 
However, at least since the middle ages where Arabic scholars such as Al-Qalqashandi used letters and their assigned numerical values, systems such as the one described here have been used to encrypt messages. The article rethinks this method with the ingredients of the digital age and recontextualises it to a quite digital world with a lot of more types of texts and documents. 
As an encipherment method, the herepresented one is rather a play, since one would not invest implementing and using an easier-to-crack method than what is already in use and can be installed out-of-the-box. The only advantage could be that the method is relatively unknown and easy to implement abd that its result mimicks frequently encountered types of noise.





\end{document}